\documentclass[12pt]{article}

\usepackage[letterpaper,hmargin=1in,vmargin=1in]{geometry}

\usepackage{graphicx,amsmath,amsfonts,amssymb}

\usepackage{cite}

\def\be{\begin{equation}}
\def\ee{\end{equation}}
\def\ba{\begin{eqnarray}}
\def\ea{\end{eqnarray}}
\def\bea{\begin{eqnarray}}
\def\eea{\end{eqnarray}}
\def\ge{\mathrel{\raise.3ex\hbox{$>$\kern-.75em\lower1ex\hbox{$\sim$}}}}
\def\la{\mathrel{\raise.3ex\hbox{$<$\kern-.75em\lower1ex\hbox{$\sim$}}}}

\def\theequation{\thesection.\arabic{equation}}
\def\simgt{\mathrel{\raise.3ex\hbox{$>$\kern-.75em\lower1ex\hbox{$\sim$}}}}
\def\simlt{\mathrel{\raise.3ex\hbox{$<$\kern-.75em\lower1ex\hbox{$\sim$}}}}

\newcommand{\bi}[1]{\bibitem{#1}}

\newcommand{\nc}{\newcommand}

\nc{\gone}{\bar g_{\pi NN}^{(1)}}
\nc{\gzero}{\bar g_{\pi NN}^{(0)}}
\nc{\al}{\alpha}
\nc{\ga}{\gamma}
\nc{\de}{\delta}
\nc{\ep}{\epsilon}
\nc{\ze}{\zeta}
\nc{\et}{\eta}

\nc{\Th}{\Theta}
\nc{\ka}{\kappa}
\renewcommand{\la}{\lambda}
\nc{\rh}{\rho}
\nc{\si}{\sigma}
\nc{\ta}{\tau}
\nc{\up}{\upsilon}
\nc{\ph}{\phi}
\nc{\ch}{\chi}
\nc{\ps}{\psi}
\nc{\om}{\omega}
\nc{\Ga}{\Gamma}
\nc{\De}{\Delta}
\nc{\La}{\Lambda}
\nc{\Si}{\Sigma}
\nc{\Up}{\Upsilon}
\nc{\Ph}{\Phi}
\nc{\Ps}{\Psi}
\nc{\Om}{\Omega}
\nc{\ptl}{\partial}
\nc{\del}{\nabla}
\nc{\ov}{\overline}
\nc{\newcaption}[1]{\centerline{\parbox{15cm}{\caption{#1}}}}
\nc{\none}{${\cal N}\!\!=\!1\;$}
\nc{\ntwo}{${\cal N}\!\!=\!2\;$}
\nc{\nfour}{${\cal N}\!\!=\!4\;$}
\nc{\nones}{${\cal N}\!\!=\!1^*\;$}
\nc{\ntwos}{${\cal N}\!\!=\!2^*\;$}
\nc{\nfoura}{${\cal N}\!\!=\!4a\;$}
\nc{\nfourb}{${\cal N}\!\!=\!4b\;$}

\nc{\slz}{SL$(2,\Z)$\;}

\def\Z{{\mathbb{Z}}}

\def\Id{\hbox{1\kern-.23em{\rm l}}}

\begin{document}

\begin{titlepage}

\rightline{FTPI--MINN--08/27}
\rightline{UMN--TH--2707/08}
\rightline{CERN-PH-TH/2008-148}
\rightline{arXiv:0807.2419}

\setcounter{page}{1}

\vspace*{0.5in}

\begin{center}

\Large{\bf Dyon dynamics near marginal stability \\ and non-BPS states}

\vspace*{1.5cm}
\normalsize

\begin{center}
{\bf Adam Ritz\boldmath{$^{\,a}$} and
  Arkady Vainshtein\boldmath{$^{\,b,\,c}$}}\\
\vspace{5mm}
{\em $^a$Department of Physics and Astronomy, 
University of Victoria, \\ Victoria BC, V8P 5C2, Canada\\\vspace{1mm}
$^b$William~I.~Fine Theoretical Physics Institute,
University of Minnesota, \\ 
Minneapolis, MN 55455, USA}\\[1mm]
{\em $^{c}$Theory Group, Physics Department, CERN, CH-1211 Geneva 23,
  Switzerland}
\vspace{3mm}
\end{center}

\smallskip
\medskip

\smallskip
\end{center}
\vskip0.4in

\centerline{\large\bf Abstract}

We derive the general form of the moduli-space effective action for the long-range interaction of two BPS 
dyons in \ntwo gauge theories. This action determines the bound state structure of various BPS and non-BPS states
near marginal stability curves, and we utilise it to compute the leading correction to the BPS-mass of zero-torsion non-BPS
bound states close to marginal stability.

\vfil
\leftline{July 2008}

\end{titlepage}

\renewcommand{\theequation}{\arabic{equation}}

\subsection*{1. Introduction}

Theories with extended \ntwo or \nfour supersymmetry in four dimensions have a BPS sector of the particle spectrum,
namely those states which preserve some fraction of the supersymmetry of the vacuum \cite{wo}. The masses, interactions and degeneracies of 
these BPS states are often exactly computable, providing a powerful window to the dynamics. One of the remarkable dynamical
principles which emerges in theories with this level of supersymmetry is electric-magnetic duality \cite{mo}, and its infinite-dimensional 
generalization $S$-duality. It was realized some time ago that the BPS spectrum then inherits a rather hierarchical structure, with the majority of the states
being viewed as bound states of a relatively small number of constituents, which are the lightest states in any charge sector
at a given point in the vacuum moduli space of the theory. This structure is quite rigid under changes of moduli with the exception
of special co-dimension one surfaces, known as curves of marginal stability (CMS), where changes occur in particular in the spectrum of
the lightest constituent states. In the context of dyonic bound states, this structure was explored in \ntwo and \nfour gauge theories 
some time ago \cite{cms} utilizing the Seiberg-Witten solution \cite{sw}, and has more recently been elucidated for 
BPS black holes in \ntwo and \nfour string theories \cite{denef00,denef,dm,cms_bh}. While much  of this dynamical structure is expected to 
extend to the more generic non-BPS sector, it is far less amenable to study as many of the powerful supersymmetric tools are no longer available.
Nonetheless, the non-BPS sector is of course of considerable interest for many reasons, not least because it provides a window into the behaviour of generic
massive states in strongly coupled gauge theories, and indeed to generic black hole states in \ntwo and \nfour string theory.

In this letter, we will consider a special class of non-BPS states which  may be
viewed as weakly bound composites of BPS constituents near curves of marginal stability. The analysis will thus be 
limited to near-CMS regions of the moduli space,
but the payoff is that explicit computations can be performed for the mass, and in principle the degeneracy, of these non-BPS states.
The interactions of BPS states are dictated by \ntwo supersymmetry, which allows us to treat the non-relativistic bound state problem
exactly even at strong coupling. This statement relies on several special features,
most importantly that  the long-range interactions of BPS constituents with charges $(n_E^{a},n_{M a})$ on 
the Coulomb branch are exactly determined by the central charges ${\cal Z} = n_E^{a} a_{a}  + n_{M a} a_{D}^{a}$ of the BPS states 
in question, and the special K\"ahler metric  on the Coulomb branch $g_{ab}$ \cite{sw,suN}. By considering
the exchange of the massless fields, the Coulomb term can be shown to take the form \cite{rv},
\be
 V_{\rm Coul}(r) = \frac{1}{4\pi r}{\rm Re}\left[ g_{ab} \frac{\ptl {\cal Z}_1}{\ptl
a_a} \frac{\ptl \bar{\cal Z}_2}{\ptl \bar{a}_b} \left(1- \frac{\bar{\cal Z}_1 {\cal Z}_2}{|{\cal Z}_1{\cal Z}_2|}\right)\right].
\label{pot1}
\ee 
This expression is particularly useful near the CMS for these two constituents, as the bound state problem simplifies to the non-relativistic level because the binding energy may be made parametrically small. 
The CMS curve(s) for the two states in question
is defined by the condition:
\be
 \left. \om\right|_{\rm CMS}=0\;\;\;\;\;{\rm where}\;\;\;\;\; e^{i\om} = \frac{\bar{\cal Z}_1 {\cal Z}_2}{|{\cal Z}_1{\cal Z}_2|}. \label{phase}
 \ee
 
In the context of supersymmetric quantum mechanics (SQM), the potential for two interacting BPS constituents with masses
$M_{i}=|{\cal Z}_{i}|$, for $i=1,2$, naturally tends to a constant as $r\to\infty$ such that the lowest eigenvalue of the SQM Hamiltonian 
vanishes for a BPS bound state with mass $M_{1+2}=|{\cal Z}_{1}+{\cal Z}_{2}|$. This implies that 
the potential is defined at large $r$ as 
\be
V(r)=\De E+V_{\rm Coul}(r) +{\cal O}(1/r^{2})\,,
\ee
in terms of the binding energy  $\De E=M_{1}+M_{2}-M_{1+2}$. 
Expansion of the potential $V(r)$ near $\om=0$ takes the form
\ba
 V(r) \!\! &=&\!\!   \frac{1}{2}\,M_r \om^2-  \frac{1}{r} \left( V_1 \om + V_2 \om^2\right) + {\cal O}(\om^3)
 +{\cal O}(1/r^{2})\nonumber\\
     \!\! &=&\!\!  \frac{1}{2}\,M_r \om^2 - \frac{1}{r} \left[ \langle Q_1,Q_2\rangle\, \om -   \frac{1}{8\pi} {\rm Re}\!\left(g_{ab} \frac{\ptl {\cal Z}_1}{\ptl
a^a} \frac{\ptl \bar{\cal Z}_2}{\ptl \bar{a}^b}\right)\om^2\right] + {\cal O}(\om^3)+{\cal O}(1/r^{2})\,,  \label{pot2}
\ea
where $M_r=M_{1}M_{2}/(M_{1}+ M_{2})$ is the reduced mass of the two BPS constituents. Close to the CMS, the sign of the Coulomb potential is determined by
the sign of the Dirac-Schwinger-Zwanziger symplectic charge product 
\be
\langle Q_1,Q_2\rangle = n_{1E}^{a} n_{2Ma}- n_{2E}^a n_{1Ma}\,, 
\ee
also known as the torsion. Thus generically as one crosses the CMS, the Coulomb potential changes sign and a bound state exists
on only one side \cite{rsvv,rv}.

This provides a simple viewpoint on the existence or otherwise of bound states and was exploited in \cite{rv} to consider the BPS spectrum
in \ntwo theories utilizing the Seiberg-Witten solution (see e.g. \cite{an} for an alternative approach). However,  this picture is incomplete as 
the relative dynamics of two BPS constituents necessarily preserves at least four supercharges in an \ntwo theory, and this structure is not 
manifest in the potential above. In general terms the structure of the relevant form of SQM has been known for some time \cite{susy}, and was studied in the 
context of BPS states by Denef \cite{denef}. We will extend this approach to incorporate the nontrivial metric for the relative translational coordinates
of the two BPS constituents, and explore how this provides a novel window on the mass corrections for non-BPS bound states. 

The general form of the worldline SQM is derived in Section~2, and shown to be consistent with the Coulomb potential in (\ref{pot2}). In Section~3, we
focus on a specific class of non-BPS bound states with zero torsion, i.e. $\langle Q_1,Q_2\rangle=0$, which are known to be BPS states
in \nfour SYM, but lift slightly from the BPS bound in \ntwo theories at weak coupling. Using the low energy description of the constituent BPS 
states we explicitly compute the non-BPS correction to the mass. We finish with some additional remarks in Section~4.

\subsection*{2. The D-term potential and the moduli-space metric}

The chiral structure of the fermionic zero modes in the background of
monopole solutions in \ntwo SYM implies that, at the classical level, the low energy dynamics 
on the moduli space of BPS dyons is realized as \nfourb SQM \cite{gps}, namely the reduction of a
chiral (0,4) supersymmetric sigma model in 1+1D, preserving four supercharges. This pairs four bosonic and four fermionic collective coordinates. 
However, while three of the bosonic coordinates reflect translations and so would be expected to survive the inclusion of generic 
quantum corrections, the fourth is compact and the quantized momentum  along this 
direction corresponds to the discrete electric charge. Thus, in a generic sector of the quantum theory where, in addition to the 
magnetic charge, the electric charge is also fixed we deal with a system having 3 bosonic and 4 fermionic
variables. Such a multiplet is in fact consistent in SQM, and arises from the reduction of a vector multiplet
in 3+1D. 

The general structure of SQM resulting from the reduction of a vector multiplet has been known for
some time \cite{susy}, and was discussed in the present context more recently by Denef \cite{denef}.
The vector multiplet decomposes to $(\vec{x},\la, D)$, comprising the coordinate vector $\vec{x}$, its chiral 
superpartner $\la$ and the auxiliary field $D$. If we consider the relative dynamics of two point-like sources, with
$\vec{x}$ the relative separation, the Lagrangian up to 2-derivative order is quite 
constrained \cite{susy,denef},
\be
 {\cal L}^{(2)} = - U D + \vec{A} \cdot \dot{\vec x} 
   + \vec{\del} U\cdot \bar{\la} \vec{\si}\la+ \frac{1}{2} \,G\hskip 0.3mm(\dot{\vec x}^{\,2} + D^2)\,,
 \ee
 where the potentials $U$, $\vec A$ and the metric $G$ are functions of the relative 
 coordinate $\vec x$.  The potentials are related by the condition
  $\vec{\del}U = \vec{\del}\!\times \!\vec{A}\,$. It follows that $U$ is a harmonic function, and requiring spherical symmetry, as is appropriate for the potential between
 two Coulomb sources, the general solution has the form $U(r) = \al - \beta/r$ with $r=|\vec x|$, constants $\al$ and $\beta$, and $\vec{A}=\beta\vec{A}^d$ with 
 $\vec{A}^d$ the unit charge Dirac monopole potential. Quantization then demands that $\beta$ be an integer.
 
 We will now specialize to the case at hand, namely the relative dynamics of two BPS states in \ntwo SYM. 
 The  2-particle interaction potential has the form
\be
 V(r) = \frac{1}{2}\,G D^2 = \frac{U^2}{2G}\,, \label{fpot}
\ee
where the above constraints on $U(r)$ should indeed hold. Since the interactions are long range, we can
also write $G = M_r + \ga/r + {\cal O}(1/r^{2})$ introducing a further constant $\ga$ and proceed to fix the constants $\al$, $\beta$ and $\ga$.
By comparing to the expansion of the Coulomb potential (\ref{pot2}) near the CMS, the binding energy immediately
fixes $\al=M_r\om$. 

To proceed, we note that for a BPS bound state to exist this system must have a supersymmetric vacuum
where $D = G^{-1}(r)U(r) =0$. Given regularity of the metric, this implies classically that $U(r)=0$ in the vacuum and the existence or
otherwise of BPS states will, generically at least, be independent of the metric $G$. Since the leading term
in the Coulomb potential (\ref{pot2}) near the CMS,  linear in $\om$, dictates the presence of BPS states it must necessarily be present in $U(r)$. 
This allows to uniquely fix  the constant $\beta$, and thus
the form of $U(r)$,
\be
 U(r) = M_r\om -  \frac{\beta}{r}\,,\qquad {\rm with}\quad\beta=\langle Q_1,Q_2\rangle\,,
 \ee
 up to corrections of ${\cal O}(\om^3)$ which are subleading near the CMS.
 
 It remains to determine the metric $G$. Indeed, it is now clear that the $1/r$ term in (\ref{fpot}) cannot
 reproduce the full Coulomb potential (\ref{pot2}) at ${\cal O}(\om^2)$ unless $\gamma\neq 0$. Classically, or in the \nfour SYM
 limit, there is enough supersymmetry to determine the form of the metric precisely. Here we will be content to
 determine the leading $1/r$ correction and, rather than extract $\ga$ from (\ref{pot2}), we will compute it using the 
 approach of Manton \cite{manton}. Near the CMS, we can consider the dynamics of the non-relativistic
 BPS point-like constituents interacting through their electric, magnetic, and scalar charges, and the metric is determined by the 
 terms quadratic in velocity.

Manton's original calculation (at the level of the leading $1/r$ terms) can be
straightforwardly repeated in the \ntwo regime using the exact quantum 
corrected charges that were computed in \cite{rv}. We begin by 
writing down the probe Lagrangian for dyon 2 in the background of
dyon 1, with both states weakly boosted,
\bea
 {\cal L}_2\! &=&\!  - \Big |{\cal Z}_2 + \de a_a \frac{ \ptl {\cal
Z}_2}{\ptl a_a}\Big |(1-\vec{v}_2^{\,2})^{1/2} \nonumber \\
 &&  + \,g_{ab}\, n_{2E}^{a} (\vec{v}_2 \cdot
\vec{A}^{\,b} - A_{0}^b) + g^{ab}\, n_{2Ma} (\vec{v}_2 \cdot
\vec{\tilde{A}}_b - \tilde{A}_{0b}).
\eea
Here $A$ and $\tilde{A}$ are the electric and dual magnetic
potentials, coupling to the state via the metric $g_{ab}$ and its
inverse, and the scalar charge is determined by the Taylor expansion
of $M_2$. When $n_E$ and $n_M$ are both nonzero we also need to
account for the Witten effect which modifies the coupling to
the electric potential.

The boosted potentials $A$ and $\tilde{A}$ induced by the 
presence of dyon 1 take the usual Li\'enard-Wiechert form
\cite{manton}, while the scalar field shift is given by \cite{rv,ps},
\be
 \de a_a = \frac{1}{4\pi r}\,g_{ab}\, \frac{
 \ptl \bar{\cal Z}_1}{\ptl \bar{a}_b} \,(1-\vec{v}_1^{\,2})^{1/2}   \,.
\ee
Expanding to quadratic order in velocities, symmetrizing to include
the dynamics of dyon 1, and dropping the free center of mass motion,
we obtain ${\cal L}_{\rm rel} = \frac{1}{2}\,G (\vec{v}_2-\vec{v}_1)^2+\cdots$, where
\be
 G= M_r + \frac{\ga}{r} + {\cal O}\Big(\frac{1}{r^2}\Big), \;\;\;\;\;\; {\rm with}\;\;\;\ga = -\frac{1}{4\pi} 
 {\rm Re}\left(g_{ab} \frac{\ptl {\cal Z}_1}{\ptl
a_a} \frac{\ptl \bar{\cal Z}_2}{\ptl \bar{a}_b}\right).
\ee
Thus, up to possible subleading terms of ${\cal O}(1/r^2)$ in the
metric, we can write down the full potential (\ref{fpot}) in the
form
\be
 V(r) = \frac{1}{2}\,\Big[M_r -\frac{1}{4\pi r} 
 {\rm Re}\Big(g_{ab} \frac{\ptl {\cal Z}_1}{\ptl
a_a} \frac{\ptl \bar{\cal Z}_2}{\ptl \bar{a}_b}\Big)\Big]^{-1}  
\Big (M_r \om - \frac{\langle
Q_1,Q_2\rangle}{r}\Big)^2 + {\cal O}(\om^3), \label{fpotres}
\ee
As a useful consistency check, by  expanding to ${\cal O}(1/r)$, we may now verify that the $1/r$ term 
agrees precisely up to ${\cal O}(\om^2)$ with potential (\ref{pot2}) computed by considering massless
exchange \cite{rv}. The full formulation as a $D$-term potential in (\ref{fpotres}) makes the constraints
of supersymmetry manifest and naturally provides an extension to higher order in $1/r$ since any
terms of ${\cal O}(1/r^2)$ in the metric will be further suppressed in $V(r)$ by a factor of $\om^2$.

\subsection*{3. The fate of zero-torsion non-BPS states}

Since $2V(r_{\rm eq})= G^{-1} U^{2}(r_{\rm eq}=0)$ for supersymmetric ground states to exist classically, we see that the
metric generically plays little role; the relevant constraint is $U(r_{\rm eq})=0$, so that
\be
 r_{\rm eq} \sim \frac{\langle Q_1, Q_2\rangle}{M_r \om}, \label{req}
\ee
which we observe only has solutions on one side of the CMS where the composite BPS state exists. This simplification allows
the quantum mechanics of generic BPS states to be studied in a simplified system where the metric is ignored, i.e. we can take
$G=M_r$, a constant. This regime was considered in detail by Denef \cite{denef}, and the BPS ground state has the form
$\Psi= \ps_\al \bar{\la}^\al |0\rangle$ in terms of $\la^{\al}$, the complex spinor superpartner of $\vec{x}$. The constraint that the
supercharges annihilate the state $Q\Ps=0$ implies $(\si^i \ptl_i+i\beta \si^iA_i^d - U(r) \Id) \ps = 0$ and the angular part of the
wavefunction is expressed in terms of the monopole harmonics of Wu and Yang \cite{wy} while the size of the
multiplet scales with $\beta$ on account of the electromagnetic contribution to the total angular momentum. The radial
wavefunction scales as $\exp(\om M_r r)$ and thus is localized on only one side of the CMS, which is the quantum version
of the fact that (\ref{req}) only makes sense for one sign of $\om$. An interesting feature of the solution is a shift of the spin
by 1/2, i.e. the ground state is a spinor, which can be interpreted as due to the induced spin coupling in the Hamiltonian,
$\De V(r) =\beta \,\vec{x}\cdot\vec{S}/r^3$ where $\vec{S} = \bar\la \vec{\si} \la/2\,$.

While the spectrum of BPS states can generally be studied independently of the $1/r$ corrections to the metric $G$, these corrections 
are crucial to the formation of non-BPS composite states. Studying these states within the present framework is often difficult as
the binding energy becomes small when the full Coulomb potential (\ref{pot1}) vanishes, which corresponds to
\be
 {\rm Re}\Big(g_{ab} \,\frac{\ptl {\cal Z}_1}{\ptl a_{a}} \,\frac{\ptl \bar{\cal Z}_2}{\ptl \bar a_{b}}\Big) = \langle Q_1,Q_2\rangle \cot \frac{\om}{2}\,,
\ee
and this condition is not generally satisfied for small $\om$. This renders the bound state problem intractable. However, 
an interesting exception that we will now focus on concerns the case where the constituents are mutually local, so the
torsion vanishes $\langle Q_1,Q_2\rangle=0$. The potential near the CMS is then provided 
purely by the metric,
\be
 V(r) = \frac{M_r\, \om^{2}}{2}\, \Big( 1 + \frac{\ga}{M_r r}\Big)^{-1} \,,
 \label{Vt}
\ee
which (for positive $\ga$) vanishes linearly at $r=0$, and tends (like $1/r$) to $M_r\om^2/2$ at
infinity. 

In \nfour SYM, zero torsion states often have a special status as they may be required by $S$-duality to exist
as bound states in the BPS spectrum. The simplest example arises for gauge group SU(3), where
the existence of a $W$ boson with charges aligned along both U(1)'s implies, via duality, 
a bound state of two distinct monopoles with unit magnetic charges along the two simple roots. The existence
of this $\{1,1\}$ state has been verified at weak coupling in various regions of the moduli space \cite{gl,lwy,sy}, starting first
with the simplest case of aligned vevs where it exists as a bound state at threshold. However, 
since a 1/2-BPS multiplet in \nfour has the same size as a generic non-BPS multiplet in \ntwo\!, we would expect that on breaking
to \ntwo\! SYM the mass
perturbation will lift the state from the BPS spectrum, and indeed this
conclusion is borne out in explicit calculations at weak coupling \cite{g99,sy}. In the present context, we cannot
draw any direct conclusions on the existence of the threshold state, since $\om$ takes the form
\be
 \om \sim {\rm Im}(\bar{a}_1 a_2) \frac{|\ta|^2}{M_1 \, M_2} +{\cal O}(\om^2)\,,
\ee
and thus vanishes for aligned vevs, as does the potential (\ref{Vt}). However, the $\{1,1\}$ state is also required by duality to exist in the
\nfour theory for misaligned vevs and indeed it duly appears in the weak coupling analysis \cite{sy}. This
state is no longer at threshold and on perturbing to \ntwo SYM  should again lift from the BPS spectrum \cite{g99}.
We can verify this in a rather straightforward manner using SQM with the potential (\ref{Vt}) and indeed we 
can also  compute the mass shift from the BPS bound.

For $\beta=0$, the system is purely bosonic and inspection of the potential (\ref{Vt}) suggests 
that we can compute the non-BPS correction to the mass as the zero-point energy. The
classical Hamiltonian has the form ${\cal H} = ({\vec p}^{\,2} + M_r^2\om^2)/(2G)$, and the Schr\"odinger 
equation, after accounting for the metric and proper ordering, can be re-expressed as an analogue Coulomb system \cite{ll,zwanziger},
\be
 \left[ \frac{{\vec p}^{\,2}}{2 M_r} - \frac{\ga E}{M_r r} 
+\frac{M_r \om^2}{2} \right]\Psi = E\, \Psi,
\ee 
with $E=M_{\{1,1\}}-M_{1+2}$ where $M_{\{1,1\}}$ is the mass of the bound state and $M_{1+2}=|{\cal Z}_{1}+{\cal Z}_{2}|$ is the BPS lower bound. The Hamiltonian in this equation
is hermitian with respect to the usual
Euclidean metric, and thus
the problem reduces to a standard hydrogenic analysis \cite{ll,zwanziger}. However, the spectrum is of course not Coulomb-like, and the scaling of the 
ground state energy is determined by the parameter $\ep = 1/\om\ga$ which we keep fixed in the near-CMS regime 
where $\om\rightarrow 0$. It will be enough to focus on the weak-coupling limit $e^2 \rightarrow 0$ and $\ga \rightarrow 2\pi/e^2$, 
so that $\ep \rightarrow e^2/(2\pi \om)$ and there are two regimes of interest, namely $\om \ll e^2$ and $\om \gg e^2$. Note that in 
this limit, the exact moduli-space metric is known and contains no corrections of ${\cal O}(1/r^2)$ \cite{gl,lwy,g99}.

\subsubsection*{3.1 Ground state for \boldmath{$\ep \gg 1$}}

In the regime $\om \ll e^2 \ll 1$, the potential becomes approximately 
Coulomb-like, $V(r) \sim - \al_{\rm eff}/r$ with $\al_{\rm eff} = \ga\om^2/2$, and the binding energy $E_{\rm bind} = M_{1}+M_{2}-M_{\{1,1\}}=\De E - E$ of the non-BPS state is
\be
 E_{\rm bind}(\ep\gg 1) \approx \frac{\De E}{4\ep^2}  \approx  \frac{\pi^2}{2}\,\frac{M_r \om^4}{e^4}\,, \label{Eb1}
\ee
where we have expressed the result in terms of the binding energy of the putative BPS state with the same charges,
$\De E=M_{1}+M_{2}-M_{1+2}\approx M_r \om^2/2$ near the CMS. We observe that for $\ep\gg 1$ the binding energy is much smaller than $\De E$ as
expected for a state lying above the BPS bound. Furthermore, we can also determine that 
 the charge radius of the state is $\langle r\rangle \sim 2/(\ga\om^2M_r)$ which can be made parametrically large by moving near the CMS for
 any finite value of $\ep$ and so this description of the non-BPS state should be reliable.

\subsubsection*{3.2 Ground state for \boldmath{$\ep\ll 1$}}

The opposite regime with $e^2 \ll \om \ll 1$ is also of interest as it corresponds to parametrically weak coupling and the 
non-BPS shift of the binding energy is better understood as a small correction $\de M$ to the mass $M_{\{1,1\}}$ of the would-be BPS $\{1,1\}$ state. 
i.e. we have
\be
 E_{\rm bind}(\ep \ll 1) \approx  \De E\left(1-2\ep\right) \approx \frac{M_r \om^2}{2}\Big( 1 - \frac{e^2}{\pi\om}\Big),
  \label{Eb2}
 \ee
verifying that the state indeed lies above the BPS bound in the \ntwo spectrum, but that for $\ep \ll 1$ it it still well-bound
at weak coupling as $\de M \ll \De E$. Note that the charge radius in this limit is given by $\langle r \rangle \sim 1/(\om M_r)$
and is again parametrically large near the CMS, justifying the low energy description of the state.

At weak coupling, $\om$ is now
given purely via a misalignment in the two vevs and thus is independent of the coupling. In this regime, 
since $\ga= 2\pi/e^2$, the fractional energy shift of this state from the BPS bound is of ${\cal O}(1)$, and thus scales as a small 
${\cal O}(e^2)$ contribution relative to the classical mass $M_r \sim 1/e^2$. More precisely, the BPS mass of the $\{1,1\}$ state is 
given by $M^{\rm BPS}_{(1,1)} = |{\cal Z}_{(1,1)}|= (4\pi /e^2)|a_1+a_2|$ where $a_i =\sqrt{2}\langle \ph \cdot \beta_i\rangle$ are the 
two Cartan vevs projected along simple roots which parametrize the moduli space at weak coupling. The true mass of
the non-BPS state is then $M_{(1,1)}^{\rm BPS} + \de M$ with 
\be
 \de M = \frac{M_r\om}{2\ga}  \approx \frac{{\rm Im}(\bar{a}_1 a_2)}{|a_1|+|a_2|} 
 = \frac{e^2}{4\pi}\frac{{\rm Im}(\bar{\cal Z}_{(1,0)} {\cal Z}_{(0,1)})}{M_{(1,1)}^{\rm BPS}}, \label{delM}
\ee
where the latter equality holds near the CMS, since $|a_1+a_2|$ differs from $|a_1|+|a_2|$ only by terms of ${\cal O}(\om^2)$,

The results in (\ref{Eb1}) and (\ref{Eb2}) provide explicit computations of the non-BPS correction to the mass for
this class of zero torsion states in the near CMS regime. It would clearly be interesting to find a deeper understanding
of these formulae and we will make some additional remarks in the following section.

\subsection*{4. Concluding remarks}

In this note, we have presented the general form of the two-body moduli-space dynamics of constituent BPS states in \ntwo supersymmetric 
theories up to two-derivative order. The results of Section~2 apply at generic strongly-coupled regions of the moduli space, with the restriction
that one considers interactions of the two constituent states near the CMS. This formulation makes supersymmetry manifest and
shows that the potential arises from a $D$-term, but also incorporates nontrivial corrections from the metric. We focused on the
latter aspect to deduce mass corrections to non-BPS bound states satisfying $\langle Q_1,Q_2\rangle =0$, but there are a couple of
other technical issues that are worthy of further comment:

\begin{itemize}
\item {\it Field-theoretic interpretation of the non-BPS mass shift}: 
In the last section, we observed that the leading shift away from
the BPS bound for the $\{1,1\}$ state for $\ep \ll 1$ took the form (\ref{delM}). It is tempting to speculate on a field theoretic interpretation
of this mass correction in a formulation where we treat the BPS states as fundamental fields, and it would be interesting to explore this in more detail. 
However, we emphasize that the normalization of the propagators means that the apparent ${\cal O}(e^2)$ shift in
(\ref{delM}) is somewhat  illusory. Inspection of the potential (\ref{Vt}) shows that at weak coupling the dependence on $e^2$ scales out and
so in terms of massless exchange it apparently corresponds to a nontrivial sum of ladder diagrams.

\item {\it Near-CMS moduli}:
Since interactions between the BPS constituents are weak near the CMS, it
is natural to ask if there is a formulation of the relative dynamics in terms
of an extended moduli space, i.e. whether the interaction potential can
be re-expressed in kinetic form as is possible at the classical level, where momentum along an additional $S^1$ 
modulus $\ch$ is identified with the relative electric charge. This reformulation also 
produces a more standard multiplet structure, with a hyper-K\"ahler target space.
We simply note here that  a naive attempt to repeat this reformulation near the 
CMS fails at any finite coupling, as the existence of one small parameter,
namely $\om$, is not sufficient to render all potentials small, i.e. ${\cal O}({\rm velocity}^2)$. While there may be various
physical reasons to anticipate that such a reformulation is not possible, e.g. the fact that we do not expect a tower of integrally
charged states at strong coupling, technically the problem first arises from the fact 
that $(a_D/a) -\ta \neq 0$ already at 1-loop, which is tied to rewriting the dynamics with a matched number
of bosonic and fermionic zero modes.

\end{itemize}

We will finish with a couple of further applications that may be interesting to explore. One concerns the fate of $S$-duality in \ntwos theories.
The soft UV nature of mass deformations breaking \nfour SUSY suggests that the induced
breaking of $S$-duality should be spontaneous, and 
thus constrained by the modular weights of the mass deformation parameters. This expectation
is elegantly born out in the vacuum sector of \nones SYM, where $S$-duality acts by permutation
on all the massive vacua \cite{dw,dorey}. This raises the question of whether a similar viewpoint may prove fruitful in 
the context of the BPS and non-BPS spectrum in the \ntwos theory, where we may expect a more
significant role to be played by the constraints of $S$-duality than in \ntwo SYM itself. A crucial question here is whether the mass perturbation
for the extra adjoint fields destabilises any of the fermionic zero modes which are crucial to the $S$-dual \nfour spectrum.
While many of these modes are localized with come characteristic scale, and thus should be stable, caution
is required in regard to some of the threshold bound states, e.g. the $\{1,1\}$ state for aligned vevs. It would be interesting
to explore how the spectrum is restructured by the extending the SQM system studied here to include the effects of the extra adjoint 
multiplets of the  \ntwos theory.

Finally, on a somewhat different theme, an interesting aspect of these states is  that, through the electromagnetic contribution to the angular 
momentum, the multiplet size and thus the degeneracy can become very large for large charges. Analogous states also have a realization within 
supergravity and may form black holes, a subject that has seen significant activity recently \cite{dm,cms_bh}. It would therefore 
be interesting to contrast the degeneracies of more general dyonic states in field theory with those 
which necessarily form horizons after coupling to gravity \cite{dm}. Related questions may also be posed in counting degeneracies on
both sides of the AdS/CFT correspondence \cite{index}.

\subsubsection*{Acknowledgements}
AV thanks the CERN Theory Group for hospitality and support.
This work was supported in part by NSERC, Canada (AR), and by the DOE under grant 
number DE-FG02-94ER408 (AV).

\newpage

\end{document}